# Ambipolar Insulator-to-metal Transition in Black Phosphorus by Ionic-liquid Gating

Yu Saito[1]* and Yoshihiro Iwasa[1, 2]

[1] Quantum-Phase Electronics Center (QPEC) and Department of Applied Physics,

The University of Tokyo, Tokyo 113-8656, Japan.

[2] RIKEN Center for Emergent Matter Science (CEMS), Wako 351-0198, Japan

* Corresponding author: saito@mp.t.u-tokyo.ac.jp

## ABSTRACT

We report ambipolar transport properties in black phosphorus using an electric-double-layer transistor (EDLT) configuration. The transfer curve clearly exhibits ambipolar transistor behavior with an on/off ratio of ~ $5 \times 10^3$. The band gap was determined as $\cong$ 0.35 eV from the transfer curve, and Hall-effect measurements revealed that the hole mobility was ~ 190 cm$^2$/Vs at 170 K, which is one order of magnitude larger than the electron mobility. By inducing an ultra-high carrier density of ~ $10^{14}$ cm$^{-2}$, an electric-field-induced transition from the insulating state to the metallic state was realized, due to both electron and hole doping. Our results suggest that black phosphorus will be a good candidate for the fabrication of functional devices, such as lateral *p-n* junctions and tunnel diodes, due to the intrinsic narrow band gap.

**Figure for TOC**

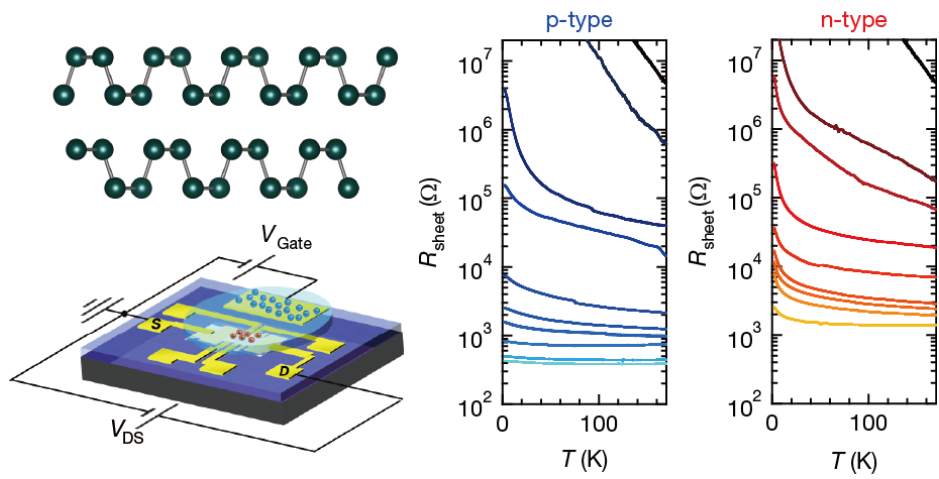

# MAIN TEXT

Since the discovery of graphene[1–3], various two-dimensional (2D) materials[4], including transition metal dichalcogenides (TMDCs), have been extensively investigated due to their promising electronic properties and potential applications. Despite the extremely high mobility in graphene, the absence of a band gap hinders the achievement of a high ON-OFF ratio. This has led to intensive research in other 2D materials that have an intrinsic band gap[4]. Among them, TMDCs, such as $MoS_2$ and $WSe_2$, which have finite band gaps, have not only enabled the fabrication of high performance field effect transistor (FET) devices[5], but have also paved the way for the realization of novel optoelectronic and valleytronic devices[6,7].

Recently, a new elemental 2D layered material, the black phosphorus (BP) monolayer, as well as its multilayers and field effect devices, have been fabricated by mechanical exfoliation[8–11]. BP is a van der Waals type semiconducting layered material with a puckered honeycomb structure, where each phosphorus atom is covalently bonded with three adjacent phosphorus atoms[12–14] (Fig. 1a), and has a direct band gap of 0.3 eV (bulk) – 2 eV (monolayer), depending on the number of layers[8]. This tunable band gap may be useful for photodetectors and other optoelectronics devices. Moreover, in addition to the bulk properties of BP, including pressure induced semiconductor to metal[15,16] and superconductor transitions[17,18], the FET of the multilayer forms has been reported by several groups to exhibit $p$-type operation[8–11], with a mobility of 200-300 $cm^2$/Vs at room temperature.

The dominant $p$-type FET action, as compared with the $n$-type, is due to unintentionally doped hole carriers in BP multilayers. However, a band structure calculation[8] has indicated that the effective masses of the conduction and the valence band do not differ very much, and a comparable FET operation might thus be possible. Examination of the electron and hole transport in a single FET configuration is crucial for understanding the carrier accumulation and transport mechanisms, as well as for fabricating functional devices based on BP. For this purpose, an electric-double-layer transistor (EDLT) using ionic liquids as the gate dielectrics is an efficient tool to widely tune the range of the Fermi energy, from the valence to the conduction band[19–21]. The electric-double-layer (EDL) formed at liquid/solid interfaces with a ~1 nm-thick Helmholtz layer allows us to effectively accumulate or deplete charge carriers. Therefore, the EDLT offers an opportunity to elucidate the nature of electron transport in BP through the realization of ambipolar transport.

**RESULTS AND DISCUSSION**

Here, we report on the EDLT operation and electric field control of the ambipolar insulator-to-metal transition in a BP thin flake. We prepared BP thin flakes from a bulk single crystal (Smart-elements, Austria) through a mechanical exfoliation method, and transferred the flakes to a $SiO_2$/Si substrate. We chose the particular flake with a thickness of 20 nm (Fig. 1b), as confirmed by atomic force microscopy, as extremely thin flakes (< 5nm) have a low mobility of ~ 10 $cm^2/Vs$[8]. We fabricated electrodes (Cr/Au, 5/70 nm) in a Hall bar configuration along with covers (Ti/$SiO_2$, 5nm/30nm) for each electrode, to prevent damaging the Au with the ionic liquid (Fig. 1b). A droplet of ionic liquid covered both the channel area and the gate electrode. The ionic liquid N,N-diethyl-N-(2-methoxyethyl)-N- methylammonium bis (trifluoromethylsulphonyl) imide (DEME-TFSI) was chosen as the gate dielectric.

Figures 2a and 2b display the ambipolar transfer characteristics of the BP-EDLT after annealing under vacuum at 370 K for 30 minutes: namely, the source-drain current $I_{DS}$ and 4-probe resistance $R_{xx}$ as a function of the gate voltage $V_{Gate}$ between -2 V and 4 V at 220 K with a sweep speed of 10 mV/s. In these measurements, $I_{DS}$ was measured with a fixed source-drain voltage $V_{DS}$ of 10 mV, and exhibited reversible behavior without any sign of degradation or anomalies due to chemical reactions, in the source-drain current, or the gate current. Before annealing, the device exhibited only *p*-type transistor behavior, but, as shown in Figs. 2a and 2b, ambipolar behavior appeared after annealing. This suggests that the influence of adsorbates on BP surfaces, which work as trapping centers for electrons, was partially negated by annealing above room temperature. At $V_{Gate} = 0$ V, the BP-EDLT is in the normally ON-state with *p*-type behavior, and a negative $V_{Gate}$ therefore accumulates holes. On the other hand, a positive $V_{Gate}$ accumulates electrons while depleting holes, reaching a charge neutrality point in the channel at $V_{Gate} \sim 1.5$ V and forming a sharp $R_{xx}$ peak. A further increase in the positive $V_{Gate}$ generates an electron inversion layer. The minimum OFF current depends on the thickness, as BP flakes are normally p-doped, and in this particular device, the ON-OFF ratio was $\sim 5 \times 10^3$. From the transfer curve shown in Fig. 2b, we defined the subthreshold swing $S = \left(\frac{d \ln I_{DS}}{d V_G}\right)^{-1}$ at $V_{DS} = 10$ mV with 114 mV/dec and 118 mV/dec for the electron and hole transport, respectively. These values are much lower than those of $SiO_2$-based BP-FETs[8], and are slightly higher than the $MoS_2$ monolayer FET (74 mV/dec)[5]. Ambipolar transfer characteristics similar to those observed in our study have also been achieved using solid-gate FETs[22].

It is known that, for many substances, the band gap can be estimated from the transfer curve, in many substances owing to the large capacitance of EDLTs[23, 24]. The change in the semiconductor Fermi level position by[25],

$$e\Delta V_{Gate} = \Delta E_F + \Delta\phi = \Delta E_F + \frac{e^2 n}{C_G}, \qquad (1)$$

where $E_F$ is the material chemical potential and $\Delta\phi$ is the variation in the electrostatic potential, which corresponds to $\frac{e^2 n}{C_G}$ ($n$ is the carrier density and $C_G$ is the geometrical capacitance.). In the case of a normal FET, the second term is usually dominant as there are localized states in the band gap due to defects in the material, and because the capacitance $C_G$ of solid-state gate dielectrics is relatively small. Conversely, EDLTs have a high geometrical capacitance, such that the contribution of the second term in Eqn. (1) can be neglected. Therefore, we can consider that $e\Delta V_{\text{Gate}}$ equals to $\Delta E_F$ for EDLTs. In the present case, the threshold voltages $V_{\text{TH}}^{\text{electron}}$ and $V_{\text{TH}}^{\text{hole}}$ were determined to be ~1.88 V and ~1.53 V for the electron and hole currents, respectively, the values being extracted from the $V_{\text{Gate}}$-linear fit are shown as a blue solid line (Figure 2a, inset). The difference between the two, $\Delta V_{\text{GAP}} = V_{\text{TH}}^{\text{electron}} - V_{\text{TH}}^{\text{hole}} \cong$ 0.35 V, corresponds to the band gap of the channel material, assuming that the potential drop occurs mainly at the interface between the channel material and the ionic liquid. The obtained value of $\cong$ 0.35 eV is roughly consistent with the bulk band gap of ~ 0.3 V [26–30], as the thickness of the present BP flake is 20 nm, which is thick enough to be considered a bulk material.

In Fig. 3, we summarize the results of the Hall-effect measurements at 170 K, at which the ionic liquid is frozen. The Hall coefficient $R_H$ showed a change in its sign at the charge neutrality point at $V_{\text{Gate}}$ ~ 1.5 V, confirming the ambipolar behavior of BP. The sheet carrier density, which is directly extracted by using the relation $n_{2D} = 1/eR_H$, is plotted against $V_{\text{Gate}}$ in Fig. 3b. $n_{2D}$ exhibited a linear behavior showing typical electrostatic charge accumulation. It is noted that the capacitance is differs between the two carrier-types: 25.8 µF/Vs and 4.8 µF/Vs for the electron and hole, respectively. While this asymmetric capacitance is quite unexpected, it can still be anticipated when an ionic liquid is used, as the size of the molecular ions accumulated in the channel differs between positive and negative $V_{\text{Gate}}$. However, the change in the capacitance occurs not at $V_{Gate} = 0$ V, but at the charge neutrality point, $V_{Gate} \sim 1.5$ V, indicating that the capacitance difference cannot be attributed to the size difference between cations and anions in the ionic liquid. Indeed, a signature of anisotropic capacitance has been observed in BP solid-gated FETs[8]. According to the reported data on gate dependent $R_H$, the capacitance $C_e$ and $C_h$ for electron and hole accumulations differs by a factor of 2 ~ 5. This is comparable to our results, suggesting that this asymmetric capacitance appears irrespective of the dielectric media (ionic liquid or $SiO_2$), and could thus be due to the intrinsic properties of BP. Although we do not yet have a clear explanation for this asymmetric behavior, the asymmetric capacitance might suggest a necessity for deeper investigation into the charge accumulation mechanism in EDLTs.

Figures 3c and 3d show the sheet conductance $\sigma_{sheet}$ and the mobility $\mu_H$, respectively, as derived from the Hall-effect measurements. The hole mobility of ~ 190 $cm^2$/Vs is much higher than that for electrons at ~ 20 $cm^2$/Vs. Considering the almost identical effective mass of electrons and holes, a much lower electron mobility exists in the BP-EDLT. This indicates that electron transport is still seriously hindered by shallow trapping centers, which remain after annealing the BP-EDLT in vacuum at 370 K. Minimizing residual traps is crucial to increasing the electron mobility. It is also noted that the hole mobility is slightly lower than the results of other groups[8,9]. It is well known that the mobility in EDLTs is always lower than in conventional solid gated FETs, as the accumulated carriers themselves work as scattering points. Another cause of the low mobility is the existence of randomly distributed cations on top of the channel, which is another source of carrier scattering. The relatively low mobility reported in the present paper is therefore not an artifact, but is instead an intrinsic effect of the EDLT device. While we believe that there is much room to improve the mobility in EDLTs, we do not believe that air exposure is the dominant cause of the low mobility. This is because the total exposure time to air was less than 1 hour, which is small enough to neglect the possibility of sample degradation due to the hydrophilicity. This is especially so considering that a clear change (such as the generation of bubbles) on the surface appears only after exposure to air for 3 days, as has been reported by other groups[8,31]. Potential solutions for this lower mobility are to transfer the black phosphorus thin flake onto a h-BN thin flake or to fabricate a sandwiched structure such as h-BN/BP/h-BN, as has been reported by other groups very recently[32–34].

Figures 4a and 4b show the sheet resistance $R_{sheet}$ as a function of the temperature $T$ for applied gate voltages between -3 V to 4 V. Despite the one-order of magnitude lower Hall mobility for the electrons, the accumulated carrier density is larger for the electrons by a factor of more than two. This results in a significant reduction in resistance by more than four orders of magnitude, on both the hole and electron channels. At the same time, we achieved the metallic state defined by $dR/dT > 0$, although an increase of the resistance due to localization occurred at low temperatures. This is the first observation of ambipolar insulator-to-metal transition in BP using ionic-liquid-gating. The continuous tuning of the conductivity and carrier density on the electron side, which has never been reported, will allow us to investigate the electron transport properties as well as the possible ground state of electron doped BP. Shao *et al*. recently reported on a theoretical prediction that the superconducting state appears in electron-doped phosphorene[35], a single atomic layer of BP, above $n_{2D} = 1.3 \times 10^{14}$ cm$^{-2}$, and that a minimum $n_{2D}$ of $2 \times 10^{14}$ cm$^{-2}$ is necessary for realizing superconductivity above $T = 2$ K, which is the lowest temperature in our experiment. Unfortunately, we did not observe superconductivity above 2 K. There are several possible reasons for this: one is simply because the electron density was not large enough for attaining superconductivity. Another possibility is that the resistance upturn observed at low temperatures occurred even at the highest gate voltage (Fig. 4b), which hinders the occurrence of superconductivity. The third possibility is that superconductivity might be restricted only in monolayer BP. Further improvement in EDLT performance may be required to achieve electric-field-induced superconductivity.

To have a further insight of the resistance upturn at low temperature, we plot the zoom-up temperature dependent conductivity below 20 K for hole and electron accumulations in Fig 5a and 5b, respectively. These figures showed contrasting behavior of hole and electron accumulation regions, namely, that the temperature-dependence for the electron accumulation is steeper than that for the hole accumulations. To understand this contrasting behavior, we analyzed its temperature dependence taking into account of the quantum correction in terms of the weal localization (WL) model which assumes not only elastic scattering but also inelastic scattering of charge carriers. This model predicts distinctive dimensionality dependence: In the case of two-dimensional (2D) systems, the contribution of WL is approximated by,

$$\Delta\sigma_{2D} = \frac{ape^2}{2\hbar}\ln T, \qquad (2)$$

where $a$ and $p$ are the constant values, $e$ is the elementary charge, $\hbar$ is the Plank constant divided by $2\pi$. In three-dimensional (3D) case, it is described by,

$$\Delta\sigma_{3D} = \frac{ce^2}{\hbar}\sqrt{\frac{\Gamma}{D}}T^{\frac{p}{2}}, \qquad (3)$$

where $c$ and $\Gamma$ are constant values, $D$ is the diffusion coefficient. If the scattering is due to the electron-phonon scattering, $p = 2$ ($T$-linear) shows the normal Fermi liquid and $p < 2$ shows the disordered system. As shown in Fig. 5a, the temperature dependent conductance for the hole accumulation is almost linear, and is well fitted by the 3D model of Eqn. (3). On the other hand, for the electron accumulated region, the steeper temperature dependence is well fitted by the 2D model of Eqn. (2). However, just above the charge neutrality point at 3 V and 2.4 V, the $\sigma_{sheet}(T)$ values are deviated from the 2D-WL fitting line and is closer to the $T$-linear line, or the expression of 3D-WL model ($p = 2$).

The above observations lead us to propose a following scenario of ambipolar charge accumulation associated with the dimensionality change from 2D to 3D nature. First, at very positive high gate voltages, where the inversion layer of electrons is formed, the quantum confinement of electrons is very strong, and the electrons are confined in the very narrow region (~ 1 nm) and the other region of the whole flake is depleted, resulting in 2D WL behavior. With decreasing the gate voltage, the thickness of accumulation layer is larger with reduced electron density, driving a 2D-3D crossover of WL for electron channel. At the charge neutrality point $V_G$ = 1.5 V, both electron and hole carriers are completely depleted. When $V_G$ is further reduced, hole doping starts, and at $V_G$ = 0 V, the whole BP flake returns back to the uniformly hole doped state, and the hole carriers behaves 3D. The enhancement of hole conductivity at $V_G$ = -3 V over $V_G$ = 0 V is only 6 times at 2 K, indicating that the bulk flake is still making a considerable contribution to the observed hole conductivity, resulting in the more 3D-like WL behavior for the hole side. The above scenario is consistent with the difference in mobility of electrons and holes shown in Fig. 3d. Since the electron carriers are strongly confined at the surface of BP, they are more subjected to surface disorder than hole carriers because the contribution from the bulk carriers are not negligible even in the hole accumulation region. This may be another case of the significantly lower mobility of electrons in addition to the shallow trapping centers discussed above.

**CONCLUSION**

In conclusion, we have presented a comprehensive view of the ambipolar transistor operation on a BP thin flake with an EDLT configuration. The band gap was determined as $\cong$ 0.35 V, which is consistent with that of the bulk material. Hall-effect measurements revealed asymmetric capacitances and differences in mobility between electrons and holes. In addition, we succeeded in inducing ultra-high carrier density and in tuning BP from a typical insulating state to a metallic state in both carrier types. Also, examination of the temperature dependence of resistance at low temperature revealed a 2D-3D crossover, when $V_G$ is scanned from electrons to holes. The 3D-like localization behavior is ascribed to the contribution from the hole carriers which is unintentionally doped over the whole flake. These results suggest a necessity of fabrication of un-doped BP flake to observed intrinsic 2D behavior of gate-induced carriers. The present work provides useful information for fabrication of BP based electronics devices for the fabrication of functional devices, such as lateral *p-n* junctions and tunnel diodes, due to the intrinsic narrow band gap

## METHOD

### Device fabrication

Black phosphorus thin flakes obtained from Smart-elements (http:// www.smart-elements.com/) were mechanically exfoliated on 300 nm thick $SiO_2$ layers on highly doped Si substrates. Since black phosphorus thin flakes are highly unstable in air, immediately after the exfoliation, we covered a resist (ZEP 520 A). The ZEP was coated onto the $SiO_2$/Si substrate by using a programmable spin-coater. Two steps spinning were performed: 500 rpm for the initial 3 s after ZEP dropping, continued by 4000 rpm for the subsequent 50 seconds. The coated substrate was heated for 3 minutes on a covered hot-plate at 423 K. Next electron- beam lithography was used to pattern the source/drain con- tact electrodes (see Main Text). We removed the resist which remained on the substrate by immersing the substrate in the N-Methyl-2-Pyrrolidone (NMP) for 40 - 60 minutes at 323 K, then spraying Acetone and immersing in the IPA (isopropyl alcohol).

### Transport measurements

All the transport measurements were performed in Physical Property Measurement System (PPMS, Quantum Design, Inc.) in a 2-300 K temperature range up to 9 Tesla under He-purged conditions. In the cooling effect measurement, when we cooled down and warmed up with the sweeping rate of 1 K/min, we maintained the rate in the whole temperature scan. In the measurement of gating effect, we first cooled down from 300 K to 220 K without gating, applied the gate voltage at 220 K, cooled down and warmed up between 220 K and 2 K, and released the gate voltage at 240 K. All the gating experiment was performed with the scanning rate of 1 K/min.

**AUTHOR INFORMATION**

**Corresponding Author**

*Email: saito@mp.t.u-tokyo.ac.jp

**Notes**

Conflict of Interest: The authors declare no competing financial interests.

**ACKNOWLEDGEMENTS**

This work was supported by the Strategic International Collaborative Research Program (SICORP-LEMSUPAR) of the Japan Science and Technology Agency, Grant-in-Aid for Specially Promoted Research (No. 25000003) from JSPS

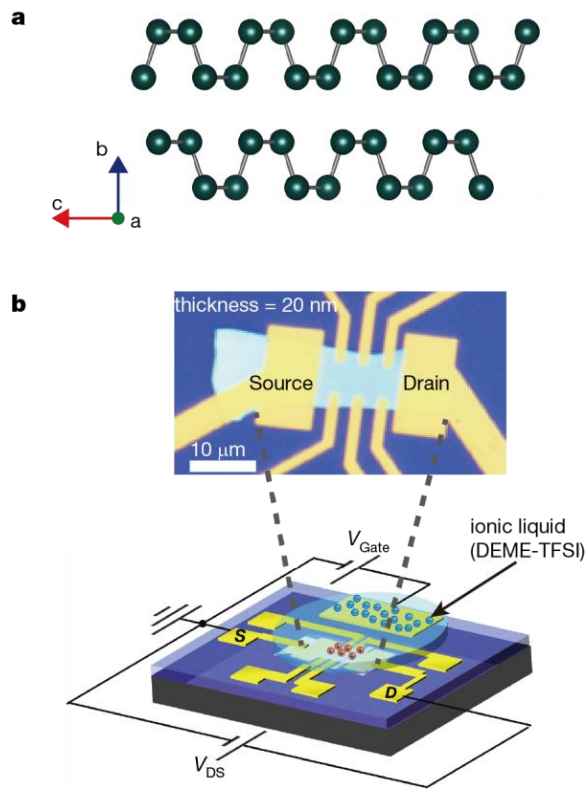

**Figure 1.** Crystal structure of black phosphorus (BP) and a schematic image of a BP-EDLT. (a) Ball-and-stick model of a BP single crystal. (b) Optical micrograph of a BP thin flake patterned with metal electrodes in a Hall bar configuration (top) and a schematic image of an EDLT device structure (bottom).

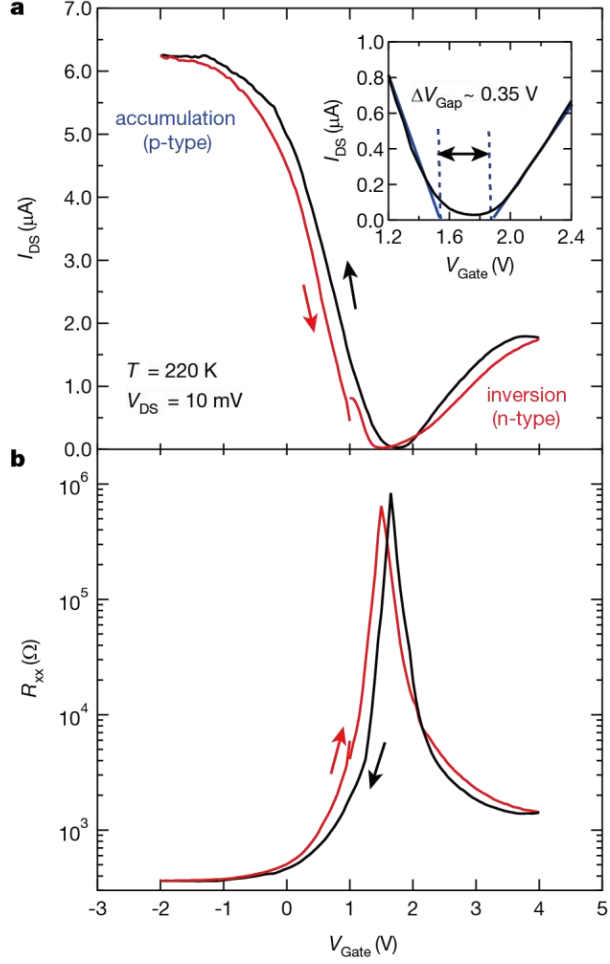

**Figure 2.** Transfer curve of a BP-EDLT. (a) Source-drain current $I_{DS}$ as a function of the gate voltage $V_{Gate}$ at $V_{DS}$ = 10 mV, showing an ambipolar operation. The inset shows a magnified image of the threshold voltage. We can extract $\Delta V_{GAP} = V_{TH}^{electron} - V_{TH}^{hole} \cong 0.35$ V by using the $V_{Gate}$-linear fit represented by the blue solid line. (b) Change in resistance $R_{xx}$ (4- probe) as a function of $V_{Gate}$.

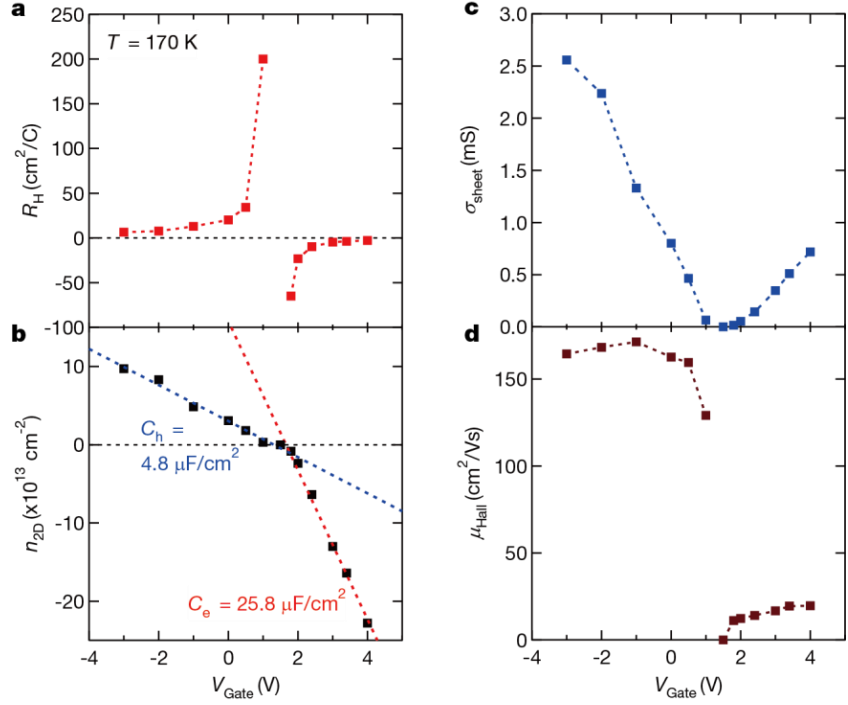

**Figure 3.** Field-effect modulation of the transport properties of the BP-EDLT. (a) The Hall coefficient $R_H$, (b) the sheet carrier density $n_{2D}$, (c) the sheet conductivity $\sigma_{sheet}$, and (d) the Hall mobility $\mu_H$ of the BP-EDLT, as a function of $V_{Gate}$ at 170 K.

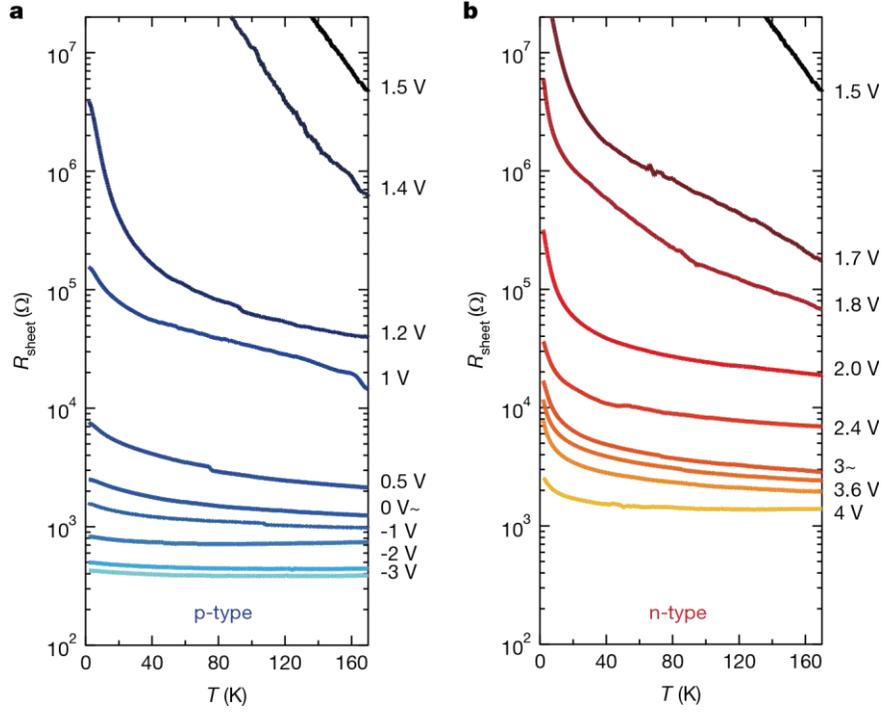

**Figure 4.** Ambipolar insulator-to-metal transition driven by an electric field. Sheet resistance $R_{sheet}$ for (a) hole transport and (b) electron transport as a function of temperature $T$, plotted on a semi-logarithmic scale, for different gate voltages $V_{Gate}$ varying from -3 V to 4V.

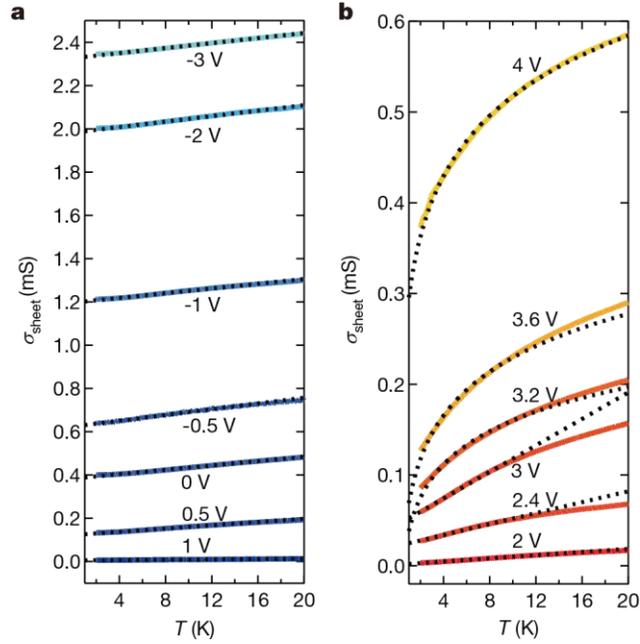

**Figure 5.** Temperature dependence of the sheet conductance $\sigma_{\text{sheet}}$ for (a) hole transport and (b) electron transport as a function of temperature $T$, for different gate voltages $V_{\text{Gate}}$. The dashed lines show 2D-WL model, and the dashed curve lines show 3D-WL model.